\documentclass[aps,prl,floatfix,twocolumn]{revtex4}
\usepackage{amsmath}
\usepackage{graphicx}

\newcommand{\dee}{\mathrm{d}}
\newcommand{\ket}[1]{|#1\rangle}
\newcommand{\bra}[1]{\langle #1|}
\newcommand{\bracket}[2]{\langle #1|#2\rangle}
\newcommand{\proj}[1]{|#1\rangle\langle #1|}

\begin{document}

\title{Coherent ultrafast measurement of time-bin encoded photons}

\author{John M. Donohue}
\email[]{jdonohue@uwaterloo.ca}
\affiliation{Institute for Quantum Computing and Department of Physics \&
Astronomy, University of Waterloo, Waterloo, Canada, N2L 3G1}

\author{Megan Agnew}
\affiliation{Institute for Quantum Computing and Department of Physics \&
Astronomy, University of Waterloo, Waterloo, Canada, N2L 3G1}

\author{Jonathan Lavoie}
\affiliation{Institute for Quantum Computing and Department of Physics \&
Astronomy, University of Waterloo, Waterloo, Canada, N2L 3G1}

\author{Kevin J. Resch}
\affiliation{Institute for Quantum Computing and Department of Physics \&
Astronomy, University of Waterloo, Waterloo, Canada, N2L 3G1}

\maketitle

\textit{Abstract} - Time-bin encoding is a robust form of optical quantum
information, especially for transmission in optical fibers. To
read out the information, the separation of the time bins must
be larger than the detector time resolution, typically on the
order of nanoseconds for photon counters. In the present work,
we demonstrate a technique using a nonlinear interaction
between chirped entangled time-bin photons and shaped laser
pulses to perform projective measurements on arbitrary time-bin
states with picosecond-scale separations. We demonstrate a
tomographically-complete set of time-bin qubit projective
measurements and show the fidelity of operations is
sufficiently high to violate the CHSH-Bell inequality by more than 6
standard deviations.

\section{Introduction}

Qubits encoded in the time-bin degree of freedom are particularly well suited for
long-distance quantum communication and fundamental
experiments~\cite{tittel98,brendel99timebin, tittel00qcrypt,
marcikic02timebin,marcikic2004fiftykm,martin12}. Time-bin states can be prepared
using an unbalanced interferometer~\cite{franson89, franson91},
where photons may take a short path and arrive early
($\ket{e}$) or a long one and arrive late ($\ket{\ell}$) with a
time difference $\tau_{e\ell}$ greater than the photon
coherence time. Measurements of time-bin states are typically performed
with an identical interferometer (see
Fig.~\ref{concept}a). However, high-fidelity measurements
require that $\tau_{e\ell}$ be greater than the detector time
resolution, which is typically much longer than the coherence
time. Experimentally, delays on the order of nanoseconds have
been used~\cite{tittel00qcrypt,martin12}; recent
advances in photon counting technology could conceivably reduce
this delay to 30~ps~\cite{hadfield2009single}. Even faster detectors would
improve time-bin encodings, allowing a higher information density while reducing the
demands on interferometric stabilization.

Ultrafast laser pulses and nonlinear optics provide a framework
for single-photon measurement on timescales much faster than
electronics~\cite{shah1988ultrafast,dayan2004tpa}. A promising
coherent nonlinear effect for single-photon ultrafast
measurements is sum-frequency generation (SFG), a process in
which two pulses interact in a nonlinear material to produce a
third with frequency equal to the sum of the
inputs~\cite{Huang1992FreqConv,kwiat_upconversion,tanzilli2005photonic,SFGent_Ramelow_2012}.
SFG in conjunction with pulse-shaping techniques is a powerful
tool for manipulating single-photon temporal
waveforms~\cite{kielpinski,eckstein2011quantum,lavoie12comp}.

In the present work, we show how sum-frequency generation and
pulse shaping enable coherent measurements of
time-bin states with a temporal separation on the picosecond
timescale. To explicitly demonstrate the coherent aspects of our
technique, we perform a tomographically
complete set of measurements on an entangled time-bin state for state
reconstruction~\cite{deBurghTomo,takesue2009tomo,Wang12tomo}.
Furthermore, we show that our measurement proceeds with sufficiently high
fidelity to convincingly violate the CHSH-Bell
inequality~\cite{bell64, CHSH}.

\section{Theory}

\begin{figure}[t!]
  \begin{center}
   \includegraphics[width=1\columnwidth]{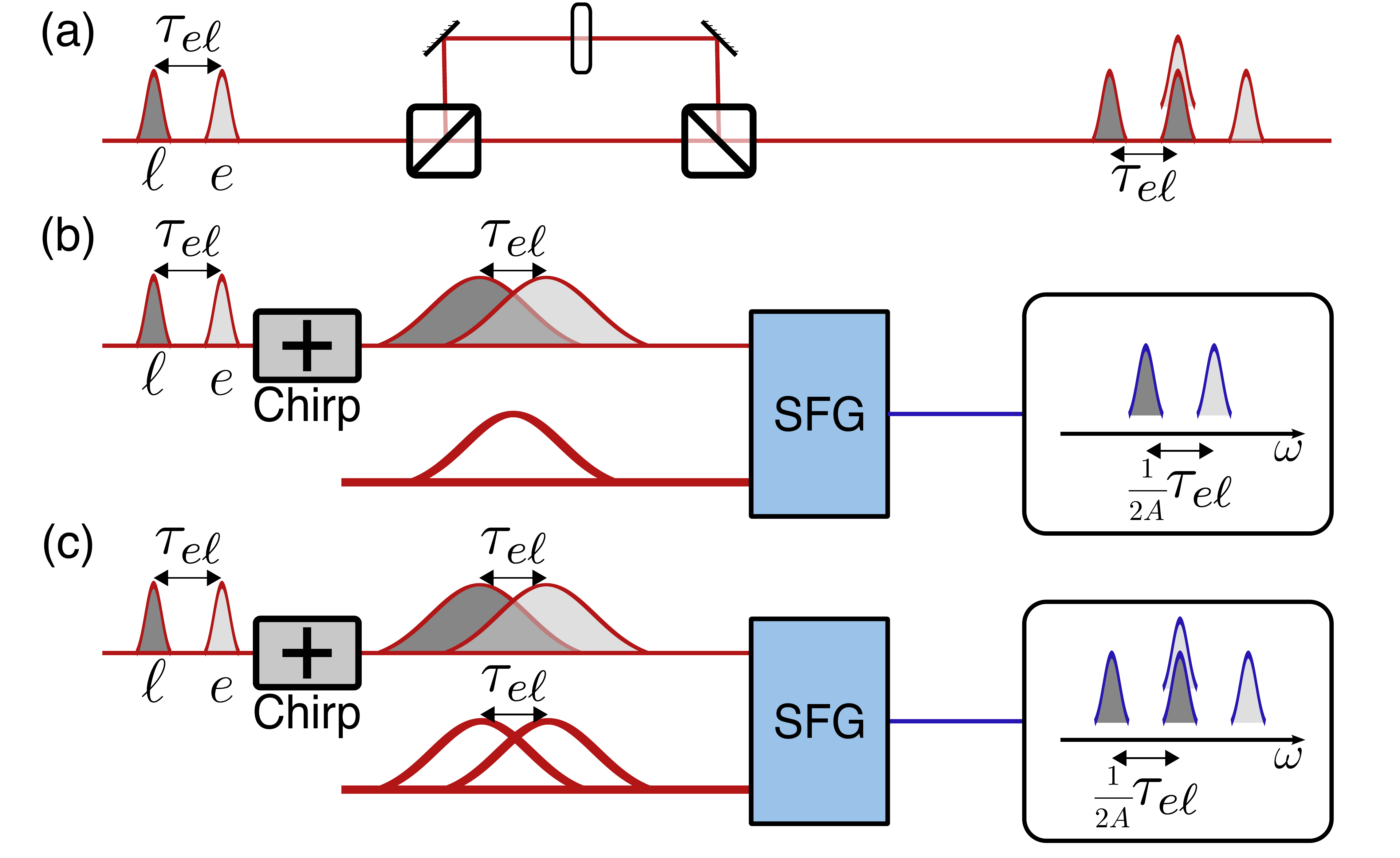}
  \end{center}
 \caption{\textbf{Measuring time-bin qubits.}
 (a) In typical time-bin measurement schemes, an input time-bin state is sent through an unbalanced interferometer matched to the
 bin separation. High-fidelity measurement requires isolating the middle output pulse,
 necessitating a large delay $\tau_{e\ell}$. (b) A photon encoding a time-bin qubit is chirped and undergoes SFG with an equal
 and oppositely chirped strong laser pulse.  The SFG contains two peaks separated in frequency
 by an amount proportional to the time delay, $\tau_{e\ell}$. (c) If the chirped strong laser
 pulse is itself in a superposition of two time bins, the output spectrum contains three peaks.  In this
 case, high-fidelity measurement requires isolating the middle frequency. The process is directly analogous to conventional
 time-bin measurement, with the signal converted from time to frequency.}\label{concept}
\end{figure}

The principle of our measurement scheme is based on SFG with
oppositely-chirped pulses. A chirped pulse is stretched such
that its instantaneous frequency varies linearly in time. By
combining two oppositely chirped pulses through SFG, the
bandwidth of the resulting pulse is drastically narrowed.
Additionally, by delaying one of the pulses, the central
frequency of the generated light changes by an amount
proportional to the delay. This has been shown for laser
pulses~\cite{BC_oppositechirp_1, BC_oppositechirp_2} and a
single photon with a strong laser pulse~\cite{lavoie12comp}. If
a pulse (or photon) is in a superposition of two time bins, it
will exit the process in a superposition of two frequencies
(see Fig.~\ref{concept}b). The process is thus a coherent
interface between time and frequency. If \emph{both} inputs are
in superpositions of time bins with the same separation, the
spectrum of the SFG output is analogous to the temporal profile
of interferometric time-bin measurement, with three distinct
frequencies. The middle peak results from the interference of
two contributions, with an intensity proportional to the
probability expected for a controllable projective measurement
(see Fig.~\ref{concept}c).

We model our scheme by expressing the electric field of a
chirped laser pulse as
\begin{equation}E(\omega;\tau,A)=f(\omega)e^{i\omega\tau}e^{iA(\omega-\omega_{\scriptscriptstyle0})^2}\end{equation},
where $\tau$ is a time delay, $A$ characterizes the chirp
strength, and
$f(\omega)=\exp[-(\omega-\omega_{\scriptscriptstyle
0})^2/(4\sigma^2)]$ is the spectral amplitude. We define a single photon
in the early time bin as $\ket{e}\propto\int\dee\omega
E(\omega;0,0)\hat{a}^\dag_{\omega}\ket{0}$ and one in the late
time bin as $\ket{\ell}\propto\int\dee\omega
E(\omega;\tau_{e\ell},0)\hat{a}^\dag_{\omega}\ket{0}$. A
time-bin qubit can be written as
\begin{equation}\ket{\psi}\approx\cos\theta\ket{e}+e^{i\phi}\sin\theta\ket{\ell}\end{equation}.
We can similarly define a superposition of two strong laser
pulses separated in time by $\tau_{e\ell}$ as
\begin{equation}E_\Lambda(\omega,\alpha,\beta)=\cos{\alpha}E(\omega;0,0)+e^{i\beta}\sin{\alpha}E(\omega;\tau_{e\ell},0),\label{SLP}\end{equation}
where $\alpha$ and $\beta$ determine the relative amplitude and
phase, respectively.

A strong laser pulse and a single photon with equal and
opposite large chirps ($A^2\sigma^4$$\gg$$1$) produce narrowband SFG with a central frequency that depends on their
relative time delay~\cite{lavoie12comp}. The SFG bandwidth is
${\sigma_3\leq1/(2\sqrt{2}A\sigma)}$, where $\sigma$ is the
smaller of the two input bandwidths. Now consider SFG between a positively chirped
time-bin qubit and a negatively chirped version of the
classical pulse from Eq.~(\ref{SLP}). For the two contributions
to the SFG from the single photon and strong laser pulse being
both early or both late, the upconverted photon will be
spectrally narrow with a central frequency $\omega_M$ equal to
the sum of the input central frequencies. Another contribution
arises from the single photon arriving early and the strong
laser pulse late, which is blue-shifted to
${\omega_B=\omega_M+\tau_{e\ell}/2A}$. Similarly, if the arrival
order is reversed, the contribution is red-shifted to
${\omega_R=\omega_M-\tau_{e\ell}/2A}$.  To spectrally separate
the three components, we require $\tau_{e\ell}$$\gg$$1/\sigma$.
Additionally, if $\tau_{e\ell}$$\ll$$1/\sigma_3$ or
equivalently $\tau_{e\ell}$$\ll$$A\sigma$, the SFG at
$\omega_M$ exhibits interference with an intensity of
\begin{equation}{I_M\propto|\cos{\theta}\cos{\alpha}+e^{i(\phi+\beta)}\sin{\theta}\sin{\alpha}|^2.}\end{equation}
This is proportional to $|\bracket{\Lambda}{\psi}|^2$, which is
the success probability of a projective measurement onto the state
${\ket{\Lambda}=\cos{\alpha}\ket{e}+e^{-i\beta}\sin{\alpha}\ket{\ell}}$,
where $\ket{\Lambda}$ is controlled by shape of the laser pulse from
Eq.~(\ref{SLP}). This technique extends naturally to arbitrary
dimensionality.  See the supplementary material for more
details.

\section{Experiment}

\begin{figure}
  \begin{center}
   \includegraphics[width=1\columnwidth]{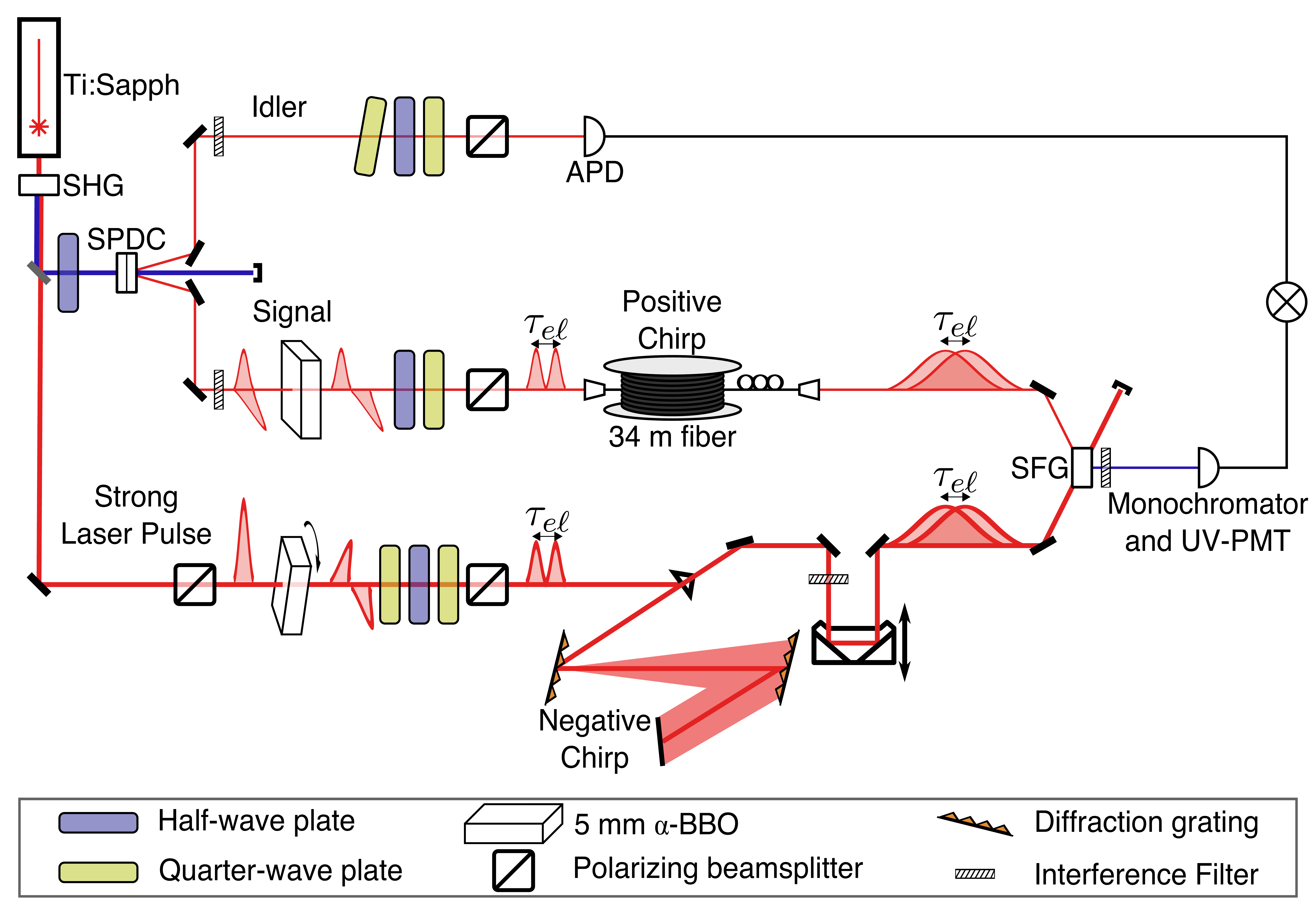}
  \end{center}
 \caption{ \textbf{Experimental setup.} Polarization-entangled photon pairs (signal and idler)
 are generated via down-conversion (SPDC) in orthogonally oriented nonlinear crystals (extra crystals used for compensation not shown). The
 signal photon is converted to a time-bin qubit using a birefringent crystal (5-mm $\alpha$-BBO) and polarizer. The signal
 acquires a positive chirp in 34~m of optical fiber. The strong
 laser pulse is prepared using an identical birefringent crystal and a
 series of waveplates to set the phase, then
 negatively chirped using gratings. The photon and laser pulse are combined in a nonlinear crystal
 to produce SFG. The middle frequency is detected using a photon counter after
 a monochromator.}\label{setup}
 \end{figure}

Our setup is shown in Fig.~\ref{setup}. A pulsed
Ti:Sapphire laser (repetition rate 80~MHz, average power 2.4~W)
centered at 790.2~nm with bandwidth 11.8~nm (FWHM) produces
0.8~W centered at 393.8~nm with a bandwidth of 1.2~nm through
frequency doubling in bismuth borate (BiBO). The UV beam is
rotated to diagonal polarization before passing through two
orthogonally-oriented $\beta$-barium borate (BBO) crystals to
produce photon pairs via type-I down-conversion (SPDC) in the
polarization state
${\ket{\Phi^+}=\frac{1}{\sqrt{2}}(\ket{HH}+\ket{VV})}$~\cite{kwiat99spdc},
where $\ket{H}$ and $\ket{V}$ are horizontal and vertical
polarizations respectively. To compensate walkoff, we inserted 1~mm of
$\alpha$-BBO into the UV beam path and 1~mm of BiBO
with a cut angle of $152.6^\circ$ into the signal
arm~\cite{lavoie2009experimental}. The signal is filtered to
810.4~nm with bandwidth $4.53$$\pm$$0.09$~nm FWHM, and the idler to
767.1~nm with bandwidth $2.37$$\pm$$0.02$~nm. We directly detect the
signal and idler photons using avalanche photodiodes (APD,
Perkin-Elmer SPCM-AQ4C). Summing the coincidence rates over all
$H/V$ combinations yields a total of 135~kHz.

We convert the signal photon from polarization to time-bin
encoding by inserting 5~mm of $\alpha$-BBO cut at $90^\circ$ into
the signal arm such that $\ket{H}$ is aligned with the
extraordinary (fast) axis and project onto diagonal
polarization with a polarizing beamsplitter to erase
polarization information, leaving the state
${\ket{\tilde{\Phi}^+}=\frac{1}{\sqrt{2}}\left(\ket{He}+\ket{V\ell}\right)}$.
The $\alpha$-BBO introduces a relative group delay of
$\tau_{e\ell}=2.16$$\pm$$0.03$~ps between the polarization
components, measured through chirped-pulse
interferometry~\cite{mazurek2013dispersion}. This delay is
greater than the photon coherence time, $1/\sigma=0.362$~ps,
fulfilling the requirements for distinct time bins.

A strong laser pulse with field
$E_\Lambda(\omega,\alpha,\beta)$ is prepared by sending the
remaining fundamental through another 5-mm $\alpha$-BBO
crystal, where rotation about the beam axis controls $\alpha$,
the relative weighting of early and late components. We can
control the phase $\beta$ between the components through the
rotation of a half-wave plate between two quarter-wave plates
set to $0^\circ$. Polarization information is then removed
using another polarizing beam-splitter. The phase $\beta$ is
four times the half-wave plate angle, with an offset due to the
birefringence in the system. This sequence simplifies
projections onto the standard states: $\ket{e}$, $\ket{\ell}$,
and $\frac{1}{\sqrt{2}}(\ket{e}+e^{i\phi}\ket{\ell})$ with
$\phi=\{-\pi/2,0,\pi/2,\pi\}$. To extend to arbitrary
projections, the rotatable $\alpha$-BBO may be replaced by a
rotatable half-wave plate and an $\alpha$-BBO set at 45
degrees.

\begin{figure}[t!]
  \begin{center}
\includegraphics[width=1\columnwidth]{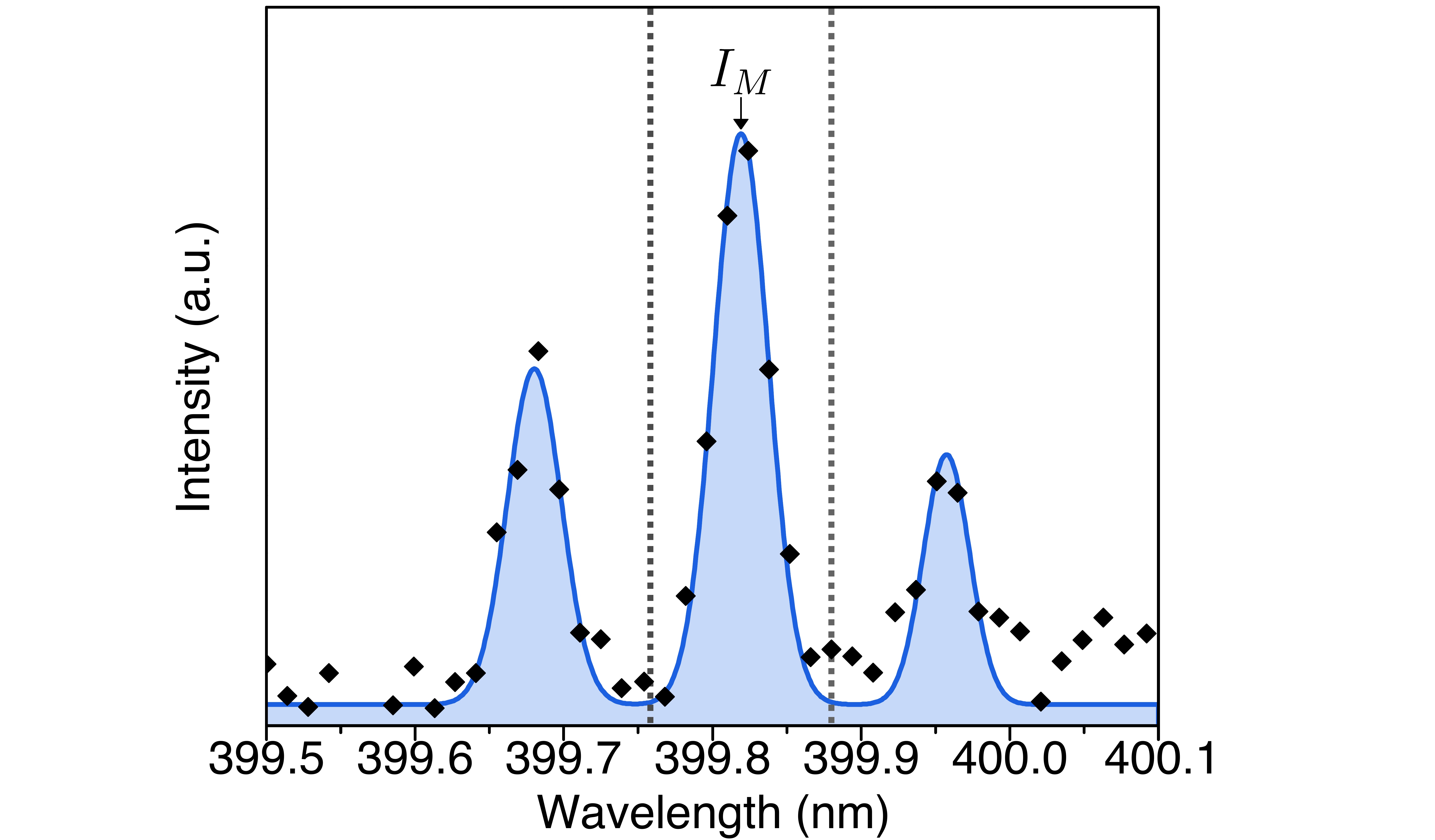}
  \end{center}
 \caption{\textbf{Sum-frequency spectrum.}  The upconverted signal spectrum
 (background subtracted) taken using our spectrometer, with $\beta$ set to $0$. A fit to the data is shown in blue.
 The monochromator selected those wavelengths that fall between the dotted lines.}\label{spec}
\end{figure}

The positive chirp of $A=(670\pm1)$$\times$$10^3~\mathrm{fs}^2$ is
applied to the single photons by passing through 34~m of
single-mode fiber. The opposite chirp on the strong laser pulse is
applied using gratings~\cite{treacy1969}. The strong laser beam
is then filtered to 785.7~nm with a bandwidth of
$11.9$$\pm$$0.3$~nm and passed through a delay line, with average
power 146~mW output. The two pulses are focused on a 1-mm BiBO
crystal phase-matched for type-I SFG, producing a UV signal
detected by photon counter (UV-PMT, Hamamatsu H10682-210).

\section{Results}

The resulting signal is sent to a fiber-coupled spectrometer
(Princeton Instruments Acton Advanced SP2750A), which we use as
either a monochromator for photon counting or a full
spectrometer. With $\beta$ set to $0$, the upconverted signal
spectrum, averaged over five 90 minute runs, is seen in
Fig.~\ref{spec} and exhibits three distinct peaks. The middle
peak, centered at $399.82$~nm, has a bandwidth of
$0.043$$\pm$$0.002$~nm. This in reasonable agreement with the
prediction of $0.035$$\pm$$0.002$~nm from the expected bandwidth
corrected for our $0.03$-nm spectrometer
resolution~\cite{lavoie12comp}. The side peaks are centered at
$399.68$~nm and $399.96$~nm. The average separation from the
main peak $\Delta\lambda_{exp}$$=$$0.138$$\pm$$0.003$~nm agrees with
the prediction $\Delta\lambda_{th}$$=$$0.137$$\pm$$0.002$~nm calculated
from the measured chirp and $\alpha$-BBO birefringence. The
separation is sufficiently large compared to the linewidth,
enabling effective filtering of the side peaks with a
monochromator window of 0.11~nm.

\begin{figure}[t!]
  \begin{center}
\includegraphics[width=1\columnwidth]{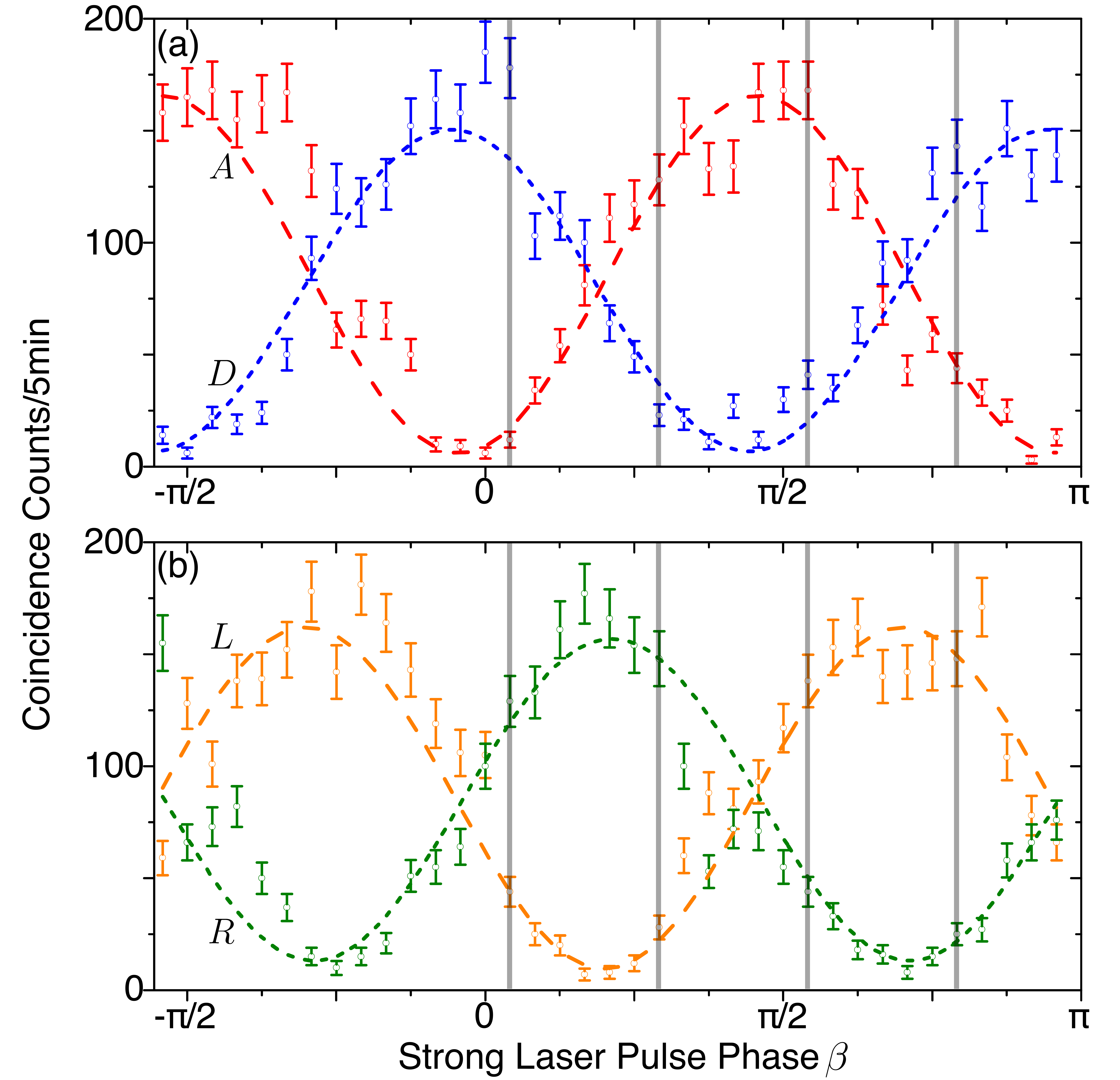}
  \end{center}
 \caption{ \textbf{Coincidence counts versus $\boldsymbol{\beta}$.} The idler is
 projected into the diagonal basis in (a) ($\ket{D}$ in blue
 and $\ket{A}$ in red) and the circular basis in (b) ($\ket{L}$
 in orange and $\ket{R}$ in green). The CHSH-Bell
 inequality was violated using the data points indicated by
 the grey lines with a value $S$$=$$2.54$$\pm$$0.08$.}\label{coin}
\end{figure}

After entangled state preparation, we vary the phase $\beta$ of
the laser pulse and record coincidences between the UV-PMT and
idler APD when the idler polarization is measured as
${\ket{D}=\frac{1}{\sqrt{2}}(\ket{H}+\ket{V})}$. We repeat this
process for idler measurements of
${\ket{A}=\frac{1}{\sqrt{2}}(\ket{H}-\ket{V})}$,
${\ket{L}=\frac{1}{\sqrt{2}}(\ket{H}+i\ket{V})}$, and
${\ket{R}=\frac{1}{\sqrt{2}}(\ket{H}-i\ket{V})}$
(Fig.~\ref{coin}). Rates of single-photon detection events were also recorded (see supplementary materials). The coincidences
oscillate sinusoidally with an average visibility among the
four curves of $89.3$$\pm$$1.7\%$. A subset of this data, for
phases indicated by vertical lines in Fig.~\ref{coin}, are
sufficient to test the CHSH-Bell inequality~\cite{bell64,
CHSH}, written as \begin{equation}{S=E(a,b)+E(a,b')+E(a',b)-E(a',b')\leq2}\end{equation}
where $E(a,b)$ is the correlation and $\{a,a',b,b'\}$ are
measurement settings. This inequality holds for local
hidden-variable models but can be violated by entangled quantum
states. We measure polarization states of the form
$\frac{1}{\sqrt{2}}(\ket{H}\pm e^{i\xi}\ket{V})$ and time-bin
states of the form $\frac{1}{\sqrt{2}}(\ket{e}\pm
e^{i\zeta}\ket{\ell})$, where the ``$+$'' and ``$-$'' outcomes
are assigned values +1 and -1, respectively. Choosing
$\xi_a$$=$$0$, $\xi_{a'}$$=$$\frac{\pi}{4}$, $\zeta_b$$=$$0.066\pi$, and
$\zeta_{b'}$$=$$0.316\pi$, the CHSH-Bell parameter was found to be
$S$$=$$2.54$$\pm$$0.08$, corresponding to a violation of the
inequality by 6.8 standard deviations.

\begin{figure}[t!!]
  \begin{center}
\includegraphics[width=1\columnwidth]{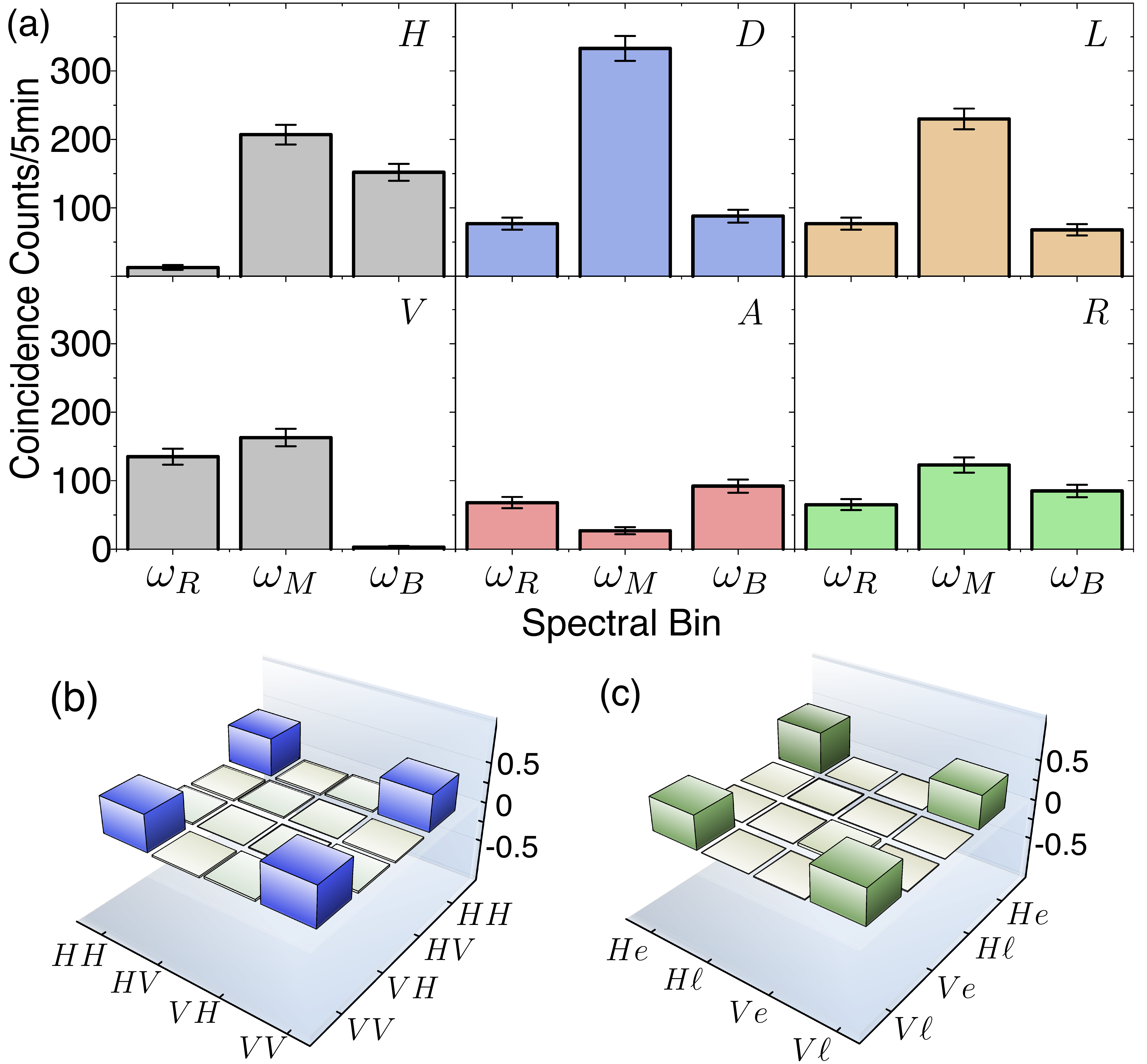}
  \end{center}
 \caption{ \textbf{Quantum state reconstruction.} (a) Coincidence counts between
   the idler and the SFG photon in each peak from Fig.~3, for $\beta$$=$$0$ and the indicated polarization measurement of the idler.  (b) Real part of the reconstructed density matrix of the initial
 two-photon polarization state produced from the SPDC source, which has a fidelity of 94\% with $\ket{{\Phi}^+}$.   c) Real part of the reconstructed density matrix
 of the polarization/time-bin state using chirped-pulse upconversion to measure the
 time-bin states; the state has 95\% fidelity with the reconstructed density matrix of the initial state.
Imaginary parts of both matrices were negligibly small and are in the supplementary material.\label{tomo}}
\end{figure}

We fixed the phase of the laser pulse to $\beta=0$ and use the
monochromator to select frequencies corresponding to the peaks
in Fig.~3. We measured the coincidence counts between the idler
for polarization measurements $\{H,V,D,A,R,L\}$ and the UV-PMT
when the monochromator was centered on each peak. The
coincidence counts for each setting and bin are shown in
Fig.~\ref{tomo}a, showing high contrast in the middle bin.
Continuing this approach for different settings of $\alpha$ and
$\beta$, we performed two-qubit tomography on our
time-bin/polarization state using an overcomplete set of 36
measurements~\cite{deBurghTomo} and iterative
maximum-likelihood reconstruction~\cite{Jezek2003tomo}.
Tomography on the initial polarization state, shown in
Fig.~\ref{tomo}b, yielded a fidelity of $94.01\pm0.02\%$ with the Bell
state $\ket{\Phi^+}$. The polarization/time-bin state
was detected at a rate of 1~Hz after upconversion and spectral
filtering, necessitating an integration time of fifteen minutes
per setting. The fidelity of the output state with the state
{$\ket{\tilde{\Phi}^+}$}
was found to be $89.4\pm0.7\%$, and the fidelity with the reconstructed
density matrix of the initial polarization state was found to
be $95.0\pm0.8\%$ (Fig.~\ref{tomo}c). Thus, our chirped-pulse
upconversion technique was able to retrieve the correlations
through quantum state tomography with minimal loss of fidelity.

\section{Conclusion}

We have demonstrated ultrafast time-bin measurements using
chirped-pulse upconversion as a coherent time-to-frequency
interface.  We showed the control necessary to perform quantum
state tomography on time-bin entangled states and sufficiently
high fidelity to convincingly violate the CHSH Bell inequality.
This technique operates at the fundamental limit for time-bin states where the
coherence time of the light, not the time resolution of the
detector, constrains the bin separation. Future work will focus
on improving the efficiency of our scheme~\cite{sensarn2009} and extensions to
time-bin qudits, which will increase the information density of
time-bin encodings.

\begin{acknowledgements}
The authors would like to thank M.D.~Mazurek, D.R.~Hamel, and
K.~Fisher for fruitful discussions. We are grateful for
financial support from NSERC, CFI, OCE, Industry Canada, and
MRI ERA.
\end{acknowledgements}

\newpage
\onecolumngrid

\newpage
\renewcommand{\thesection}{S.\arabic{section}}
\renewcommand{\thesubsection}{\thesection.\arabic{subsection}}
\makeatletter 
\def\tagform@#1{\maketag@@@{(S\ignorespaces#1\unskip\@@italiccorr)}}
\makeatother
\makeatletter
\makeatletter \renewcommand{\fnum@figure}
{\figurename~S\thefigure}
\makeatother
\renewcommand{\figurename}{Figure}
\setcounter{equation}{0}

\appendix

\newpage
\section{Supplementary: Derivation of time-bin interference}

We follow the approach of \cite{lavoie12comp} by modeling the creation of upconverted single photons through the interaction Hamiltonian $H$ of a second-order nonlinear process as \begin{equation}H=\iiint\dee\omega_1\dee\omega_2\dee\omega_3\hat{a}^{(1)}_{\omega_1}\hat{a}^{(2)}_{\omega_2}\hat{a}^{\dag(3)}_{\omega_3}e^{-i(\omega_1+\omega_2-\omega_3)t}+h.c.\end{equation} We make the approximations that the input pulses are relatively narrowband, phasematching is perfect, only one photon exists in the system at a time, the second-mode input is strong (replace $\hat{a}_{\omega_2}$ with complex constant $\alpha_{\omega_2}$), and no frequency correlations exist between the signal and idler (for a treatment of the frequency-correlated case, see the supplementary material of \cite{lavoie12comp}).

An arbitrary single-photon time-bin state is defined in the main text as $\ket{\psi}\approx\cos{\theta}\ket{e}+e^{i\phi}\sin{\theta}\ket{\ell}$, where $\ket{e}$ and $\ket{\ell}$ respectively define a single photon in the early and late time bin. Thus we write a chirped time-bin state with bin separation $\tau_{e\ell}$ as $\ket{\psi}=\int\dee\omega_1E_1(\omega_1)\hat{a}^\dag_{\omega_1}\ket{0}$, where \begin{equation}E_1(\omega_1)\propto e^{-\frac{(\omega_1-\omega_{01})^2}{4\sigma_1^2}}e^{iA(\omega_1-\omega_{01})^2}(\cos\theta+e^{i\omega_1\tau_{e\ell}}e^{i\phi}\sin\theta).\end{equation} The field of the correspondingly anti-chirped strong laser pulse is similarly defined as \begin{equation}E_\Lambda(\omega_2)\propto e^{-\frac{(\omega_2-\omega_{02})^2}{4\sigma_2^2}}e^{-iA(\omega_2-\omega_{02})^2}e^{i\omega_2\delta}(\cos\alpha+e^{i\omega_2\tau_{e\ell}}e^{i\beta}\sin\alpha),\end{equation} where $\beta$ and $\delta$ are constant.

The state of the upconverted photon to first-order perturbation theory is \begin{equation}\ket{\psi_f}\propto\iint\dee\omega_3\dee\omega_1 E_1(\omega_1)E_\Lambda(\omega_3-\omega_1)\hat{a}^\dag_{\omega_3}\ket{0}\end{equation} If the chirp is large ($A^2\sigma_i^4\gg1$), the spectral amplitude of the upconverted single photon is \begin{equation}\begin{array}{lll}E_3(\omega_{3})&\propto&e^{iA(\omega_3-\omega_{01}-\omega_{02})^2\frac{\sigma_1^2-\sigma_2^2}{\sigma_1^2+\sigma_2^2}} \Bigl[\cos\theta\sin\alpha\left(e^{-\frac{4A^2\sigma_1^2\sigma_2^2}{\sigma_1^2+\sigma_2^2}(\omega_3-\omega_{01}-\omega_{02}-\frac{\delta+\tau_{e\ell}}{2A})^2-\frac{(\delta+\tau_{e\ell})^2}{16A^2(\sigma_1^2+\sigma_2^2)}}e^{i(\beta+\tau_{e\ell}\omega_{02}+\frac{\omega_3-\omega_{01}-\omega_{02}}{\sigma_1^2+\sigma_2^2}\tau_{e\ell}\sigma_2^2)}\right)\\ &&+\left(\cos\theta\cos\alpha+e^{i(\beta+\phi+\tau_{e\ell}\omega_3)}\sin\theta\sin\alpha\right)\left(e^{-\frac{4A^2\sigma_1^2\sigma_2^2}{\sigma_1^2+\sigma_2^2}(\omega_3-\omega_{01}-\omega_{02}-\frac{\delta}{2A})^2-\frac{(\delta)^2}{16A^2(\sigma_1^2+\sigma_2^2)}}\right)\\
&&+\sin\theta\cos\alpha\left(e^{-\frac{4A^2\sigma_1^2\sigma_2^2}{\sigma_1^2+\sigma_2^2}(\omega_3-\omega_{01}-\omega_{02}-\frac{\delta-\tau_{e\ell}}{2A})^2-\frac{(\delta-\tau_{e\ell})^2}{16A^2(\sigma_1^2+\sigma_2^2)}}e^{i(\phi+\tau_{e\ell}\omega_{01}+\frac{\omega_3-\omega_{01}-\omega_{02}}{\sigma_1^2+\sigma_2^2}\tau_{e\ell}\sigma_1^2)}\right)\Bigr] \end{array}\end{equation} which can be simplified by grouping terms to, \begin{equation}E_3(\omega_{3})\propto e^{i\beta}\cos\theta\sin\alpha E_{e\ell}(\omega_3)+\left(\cos\theta\cos\alpha E_{ee}(\omega_3)+e^{i(\beta+\phi)}\sin\theta\sin\alpha E_{\ell\ell}(\omega_3)\right)+e^{i\phi}\sin\theta\cos\alpha E_{\ell e}(\omega_3).\end{equation}

From this expression, it can be seen that four contributions are made to the final spectral profile, each corresponding to one of four spectral peaks. The spectral intensity is proportional to the square of the amplitude, and can be written for each peak as \begin{equation}\left|E_i(\omega_3)\right|^2\propto S_i(\omega_3)=e^{-\frac{(\omega_3-\omega_{03i})^2}{2\sigma_3^2}}.\end{equation} From this, we find the RMS width of each peak to be given by \begin{equation}\sigma_3=\frac{1}{4A}\sqrt{\frac{1}{\sigma_1^2}+\frac{1}{\sigma_2^2}}.\end{equation} By setting $\sigma=\min\{\sigma_1,\sigma_2\}$, we can bound the upconverted frequency as $\sigma_3\leq1/(2\sqrt{2}A\sigma)$. Two of these peaks, $E_{ee}(\omega_3)$ and $E_{\ell\ell}(\omega_3)$, have the same central frequency $\omega_M$ and differ only by a linear phase factor $e^{i\tau_{e\ell}\omega_3}$, while $E_{e\ell}(\omega_3)$ is blue-shifted to $\omega_B$ and $E_{\ell e}(\omega_3)$ is red-shifted to $\omega_R$. The central frequency of each is given by \begin{equation}\omega_{03i}=\omega_{01}+\omega_{02}+\frac{\delta_i}{2A},\end{equation} where $\delta_{M}=\delta$, $\delta_{B}=\delta+\tau_{e\ell}$, and $\delta_{R}=\delta-\tau_{e\ell}$.

While angular frequency and bandwidths expressed in terms of RMS widths are preferable for derivations, values are generally reported in wavelengths and full-widths at half-maximum (FWHM). The central wavelength of the upconverted signal can be found as \begin{equation}\lambda_{03i}=\frac{\lambda_{01}+\lambda_{02}}{\lambda_{01}+\lambda_{02}+\frac{\delta_i}{4\pi cA}\lambda_{01}\lambda_{02}}\approx\frac{\lambda_{01}\lambda_{02}}{\lambda_{01}+\lambda_{02}}-\frac{\lambda_{01}^2\lambda_{02}^2}{4\pi cA(\lambda_{01}+\lambda_{02})^2}\delta_i\end{equation} with a bandwidth FWHM of \begin{equation}\Delta\lambda_{3}=\lambda_{03i}^2\frac{2\ln2}{4\pi cA}\sqrt{\frac{\lambda_{01}^4}{(\Delta\lambda_1)^2}+\frac{\lambda_{02}^4}{(\Delta\lambda_2)^2}}.\end{equation}

A projective measurement on a state $\ket{\psi}=\cos{\theta}\ket{e}+e^{i\phi}\sin{\theta}\ket{\ell}$ of $\ket{\chi}=\cos{\Theta}\ket{e}+e^{i\Phi}\sin{\Theta}\ket{\ell}$ has a success probably of \begin{equation}\begin{array}{rcl}|\bracket{\chi}{\psi}|^2&=&|\cos{\theta}\cos{\Theta}+e^{i(\phi-\Phi)}\sin{\theta}\sin{\Theta}|^2\\
&=&\cos^2{\theta}\cos^2{\Theta}+2\cos{(\phi-\Phi)}\cos{\theta}\cos{\Theta}\sin{\theta}\sin{\Theta}+\sin^2{\theta}\sin^2{\Theta}.\end{array}\end{equation} We define the side peaks as arising from the fields $E_{\ell e}(\omega_3)$ and $E_{e\ell}(\omega_3)$, and the middle peak as from the sum of $E_{ee}(\omega_3)$ and $E_{\ell\ell}(\omega_3)$, with central frequencies $\omega_R$, $\omega_B$, and $\omega_M$ respectively. The spectral intensity of the middle peak is \begin{eqnarray}S_M(\omega_3)&\propto&\left|\cos\theta\cos\alpha E_{ee}(\omega_3)+e^{i(\beta+\phi)}\sin\theta\sin\alpha E_{\ell\ell}(\omega_3)\right|^2\\ &\propto&\left|\cos\theta\cos\alpha+e^{i(\phi+\beta+\tau_{e\ell}\omega_3)}\sin\theta\sin\alpha\right|^2\left[e^{-\frac{8A^2\sigma_1^2\sigma_2^2}{\sigma_1^2+\sigma_2^2}(\omega_3-\omega_{01}-\omega_{02}-\frac{\delta}{2A})^2-\frac{\delta^2}{8A^2(\sigma_1^2+\sigma_2^2)}}\right],\end{eqnarray} which can then be integrated over $\omega_3$ to find a central peak intensity of \begin{equation}I_M\propto\cos^2\theta\cos^2\alpha+2e^{-\frac{(\sigma_1^2+\sigma_2^2)\tau_{e\ell}^2}{32A^2\sigma_1^2\sigma_2^2}}\cos{(\phi+\beta+\omega_{03}\tau_{e\ell})}\cos\theta\cos\alpha\sin\theta\sin\alpha+\sin^2\theta\sin^2\alpha.\end{equation} The ideal visibility of the interference (for $\alpha=\theta=\frac{\pi}{4}$) can be found as \begin{equation}V_{theo}=\frac{I^{(max)}_{M}-I^{(min)}_{M}}{S^{(max)}_{M}+S^{(min)}_{M}}=e^{-\frac{(\sigma_1^2+\sigma_2^2)\tau_{e\ell}^2}{32A^2\sigma_1^2\sigma_2^2}}=e^{-\frac{\sigma_3^2\tau_{e\ell}^2}{2}}.\end{equation} Thus, in order to exhibit highly visible interference, the time $\tau_{e\ell}$ between the bins should be much smaller than the temporal bandwidth of the output pulse, i.e. $\tau_{e\ell}\ll\frac{1}{\sigma_3}$. If $\tau_{e\ell}$ is sufficiently small, the intensity of the central peak can be approximated, \begin{equation}I_M\propto \left|\cos\theta\cos\alpha+e^{i(\phi+\beta+\tau_{e\ell}\omega_{03})}\sin\theta\sin\alpha\right|^2.\end{equation} By absorbing the constant phase factor $\tau_{e\ell}\omega_{03}$ into the preparation of the strong laser pulse phase $\beta$, the intensity of the central peak is found to be directly proportional to a projective measurement onto the state $\ket{\Lambda}=\cos{\alpha}\ket{e}+e^{-i\beta}\sin{\alpha}\ket{\ell}$ as long as the side peaks can be clearly distinguished.

In order to clearly distinguish the side peaks from the middle peak, it is important that they do not overlap in frequency (i.e. $\omega_{B}$ and $\omega_R$ more than $\sigma_3$ separated from $\omega_{M}$). The following condition must be met to ensure that the peaks are clearly separable in frequency: \begin{equation}\tau_{e\ell}>\sqrt{\frac{1}{\sigma_1^2}+\frac{1}{\sigma_2^2}},\end{equation} which is equivalent to stating that the delay between the two bins must be greater than the longer coherence time of the input pulses.

In summary, for ideal interference and filtering capability, the time delay between bins must satisfy \begin{equation}\sqrt{\frac{1}{\sigma_1^2}+\frac{1}{\sigma_2^2}}<\tau_{e\ell}<4A\sqrt{\frac{\sigma_1^2\sigma_2^2}{\sigma_1^2+\sigma_2^2}}.\end{equation} In our experiment, the lower bound is approximately $0.2$~ps and the upper bound is approximately $14$~ps. Thus, our bin separation of $2.16$~ps is well within the appropriate parameters.

\newpage
\section{Supplementary: Extension to time-bin qudits}

Time-bin encodings naturally extend to higher dimensions by simply adding more bins. To extend our measurement technique to higher dimensions, we similarly require more pulses in the classical beam. In doing so, photons in a certain time bin will upconvert to the central frequency $\omega_M$ only if they are upconverted by the strong laser pulse component with the same time delay.

For a time-bin qudit of dimension $N$, we define basis states with a time delay of $\tau$ between them as \begin{equation}\ket{t_j}\propto\int\dee\omega E(\omega)e^{ij\omega\tau}\hat{a}_{\omega}^\dag\ket{0}.\end{equation} We represent an arbitrary superposition state $\ket{\psi}$ with complex constants $c_j$ as \begin{equation}\ket{\psi}=\sum_{j=0}^{N-1}c_j\ket{t_j}\end{equation} and set $E(\omega)$ (after chirping) to be a Gaussian envelope, \begin{equation}E(\omega_1)=e^{-\frac{(\omega_1-\omega_{01})^2}{4\sigma_1^2}}e^{iA(\omega_1-\omega_{01})^2}.\end{equation} We similarly define a strong laser pulse as before to be a superposition of $N$ classical fields with complex constants $d_j$ as \begin{equation}E_\Lambda(\omega_2)= e^{-\frac{(\omega_2-\omega_{02})^2}{4\sigma_2^2}}e^{-iA(\omega_2-\omega_{02})^2}\sum_{j=0}^{N-1}d_je^{ij\omega_2\tau}.\end{equation} Note that we have neglected the constant time difference $\delta$ for simplicity.

By following the same procedure as for the two-dimensional case, we find that the upconverted pulse once again consists of numerous frequency peaks. We concentrate on the middle peak, with a field $E_M(\omega_3)$ centered on $\omega_{03}=\omega_{01}+\omega_{02}$, which arises when the $\ket{t_j}$ term of the qudit field is upconverted by the $j^{th}$ strong laser pulse. This middle field can be found as \begin{equation}E_M(\omega_3)=\int\dee\omega_1 e^{-\frac{(\omega_1-\omega_{01})^2}{4\sigma_1^2}}e^{iA(\omega_1-\omega_{01})^2} e^{-\frac{(\omega_3-\omega_1-\omega_{02})^2}{4\sigma_2^2}}e^{-iA(\omega_3-\omega_1-\omega_{02})^2} \sum_{j=0}^{N-1}c_jd_je^{ij\omega_3\tau}.\end{equation} The integrated spectral intensity of the middle peak can then be calculated as \begin{equation}I_M(\omega_3)=\int\dee\omega_3E^*_M(\omega_3)E_M(\omega_3)\propto\sum_{j=0}^{N-1}\sum_{k=0}^{N-1} e^{-(j-k)^2\frac{\sigma_3^2\tau^2}{2}}e^{i(j-k)\omega_{03}\tau}c^*_kd^*_kc_jd_j.\end{equation} The success probability of a general projective measurement of $\ket{\psi}$ onto $\ket{\chi}=\sum_{j=0}^{N-1}x_j\ket{t_j}$ can be expressed as \begin{equation}|\bracket{\chi}{\psi}|^2=\sum_{j=0}^{N-1}\sum_{k=0}^{N-1} c^*_kx_kc_jx^*_j.\end{equation} Thus, in an analogous fashion to the qubit case, if the peaks are clearly separable and $\tau\ll\frac{1}{\sigma_3}$, the intensity of the middle peak is proportional to the success probability of a projective measurement onto \begin{equation}\ket{\Lambda}=\sum_{j=0}^{N-1}d^*_je^{-ij\omega_{03}\tau}\ket{t_i}.\end{equation} Thus, the scheme generalizes to higher-dimensional time-bin states in a straightforward manner.

\newpage
\section{Supplementary: Quantum-state reconstruction details}

\begin{figure}[h!]
  \begin{center}
\includegraphics[width=0.5\columnwidth]{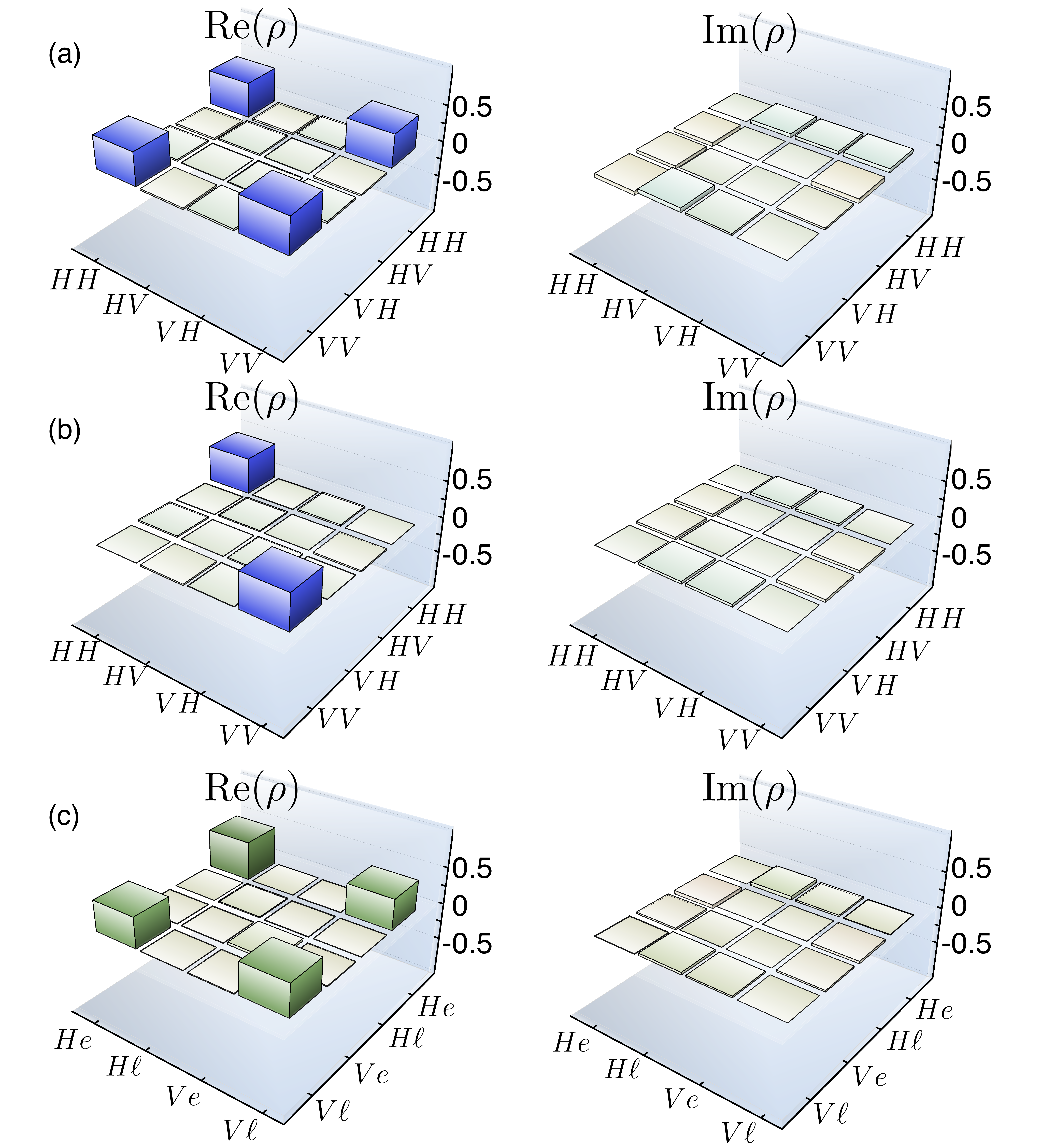}
  \end{center}
 \caption{\textbf{Full quantum-state reconstruction} (a) Tomography of the initial two-qubit polarization state. b) Tomography of the polarization state after the $\alpha$-BBO crystal has been inserted. (c) Tomography on the polarization/time-bin state using chirped-pulse upconversion to implement measurements, retrieving the correlations of the initial state.\label{tomo2}}
\end{figure}

Quantum-state reconstruction was performed using iterative maximum-likelihood tomography~\cite{Jezek2003tomo} with an overcomplete set of 36 measurement settings~\cite{deBurghTomo}. The initial polarization state was measured with an integration time of 5~s per setting and reconstructed to the density matrix of Fig.~S\ref{tomo2}a. The reconstructed density matrix has a fidelity, defined as ${\bra{\chi}\rho\ket{\chi}}$ for a pure state $\ket{\chi}$, of $0.9400\pm0.0002$ with $\ket{\Phi^+}=\frac{1}{\sqrt{2}}(\ket{HH}+\ket{VV})$. The purity of this density matrix, defined as $\textrm{Tr}\rho^2$, was found to be $0.9129\pm0.0004$. We determined errors on our fidelities and purities using Monte Carlo techniques with 400 iterations and assuming Poissonian error.

Fig.~S\ref{tomo2}b shows the reconstructed density matrix of the polarization state after the $\alpha$-BBO was inserted. This density matrix has a fidelity of $0.9683\pm0.0002$ with the classically-correlated state $(\proj{HH}+\proj{VV})/2$ and a purity of $0.4811\pm0.0002$. The entanglement in the system is no longer noticeable through polarization measurements as the time delay has separated horizontal and vertical components to outside of their coherence length.

Using our time-bin measurement technique on the signal photon instead of polarization measurements (Fig.~S\ref{tomo2}c) retrieves the correlations of the initial state. The lower measurement efficiency necessitated three loops for tomography, each with an integration time of 300~s per measurement setting. The reconstructed density matrix has a fidelity of $0.894\pm0.007$ with $\ket{\tilde{\Phi}^+}=\frac{1}{\sqrt{2}}(\ket{He}+\ket{V\ell})$ and a purity of $0.818\pm0.013$. It is more relevant, however, that the fidelity with the first reconstructed density matrix is $0.950\pm0.008$, indicating the high fidelity of operation of our measurement technique.

\newpage
\section{Supplementary: Coincidence rate vs. $\boldsymbol{\beta}$ details}

\begin{figure}[h!]
  \begin{center}
\includegraphics[width=1.0\columnwidth]{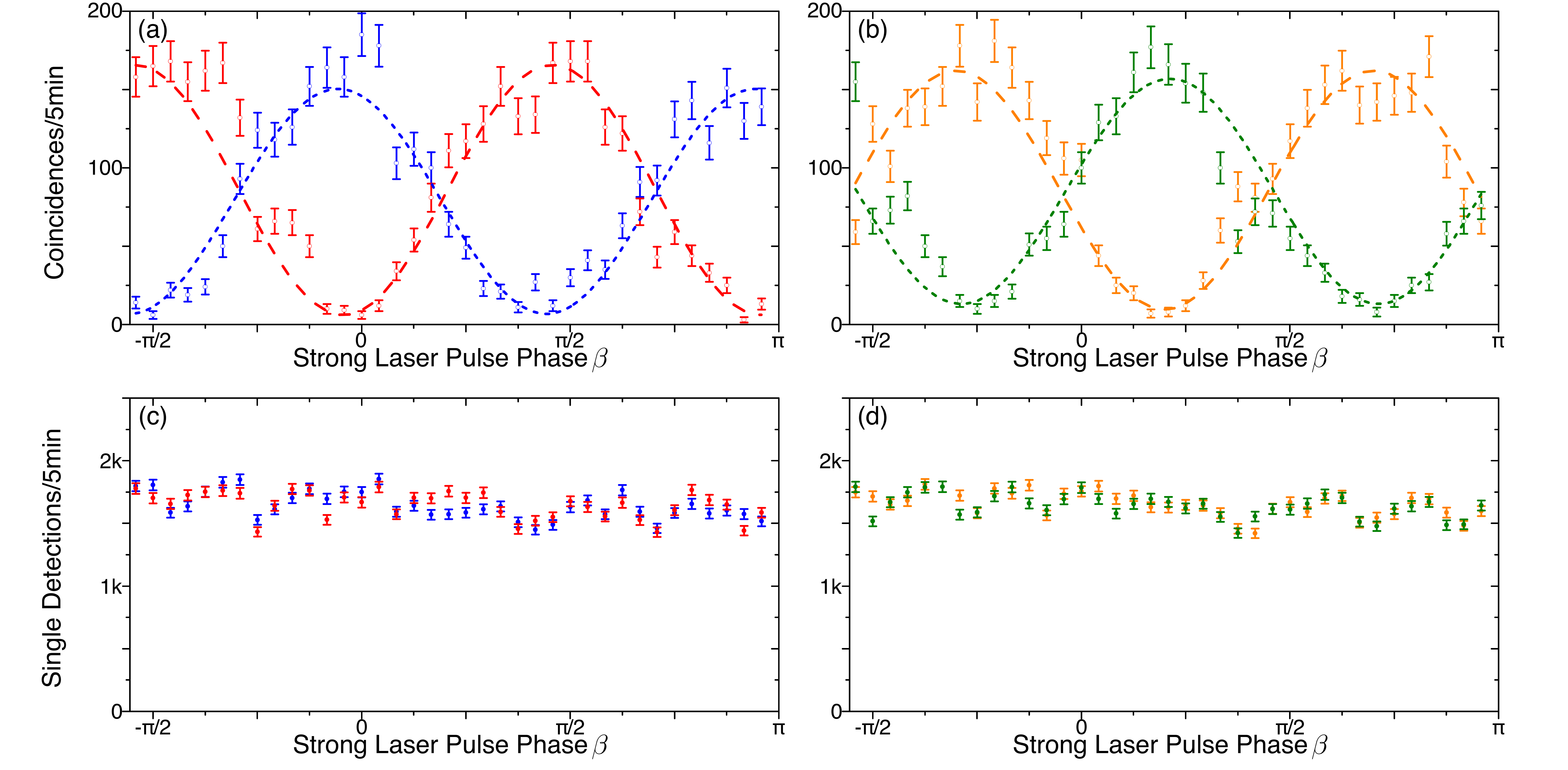}
  \end{center}
 \caption{\textbf{Full interference fringes} The coincidence rate between the polarization-encoded idler and the time-bin encoded signal is shown as the phase $\beta$ of the strong laser pulse is rotated, with the idler projected onto $\ket{D}$ (a, blue), $\ket{A}$ (a, red), $\ket{L}$ (b, orange), and $\ket{R}$ (b, green). In (c-d), the single-detection rate of the upconverted signal is shown to be nearly constant as the phase is rotated.}\label{coin2}
\end{figure}

For an entangled state of the form $\ket{\tilde{\Phi}^+}=\frac{1}{\sqrt{2}}(\ket{He}+\ket{V\ell})$, we take separable measurements on bases mutually unbiased from the H/V and $e$/$\ell$ bases, written in the form \begin{equation}\ket{M_1(\gamma)}\otimes\ket{M_2(\beta)}=\frac{1}{\sqrt{2}}(\ket{H}+e^{i\gamma}\ket{V})\otimes\frac{1}{\sqrt{2}}(\ket{e}+e^{i\beta}\ket{\ell}).\end{equation} In doing so, we expect to see a coincidence rate proportional to \begin{equation}|(\bra{M_1(\gamma)}\otimes\bra{M_2(\beta)})\ket{\tilde{\Phi}^+}|^2=\frac{1}{2}\cos^2\frac{\gamma+\beta}{2},\end{equation} which can vary between 0 and $\frac{1}{2}$ as the phases $\beta$ and $\gamma$ are altered. If we instead look at only half of the two-qubit system, we find the single-event rate for each side to be proportional to \begin{equation}\bra{M_1(\gamma)}\textrm{Tr}_2[\proj{\tilde{\Phi}^+}]\ket{M_1(\gamma)}= \bra{M_2(\beta)}\textrm{Tr}_1[\proj{\tilde{\Phi}^+}]\ket{M_2(\beta)}=\frac{1}{2}.\end{equation} Thus, when the phases are varied, we expect oscillations in the coincidence rate but a stable rate of single-event detections~\cite{franson89}.

Fig.~S\ref{coin2} shows the coincidence rate (Fig.~S\ref{coin2}(a-b)) and single-event rate (Fig.~S\ref{coin2}(c-d)) of the time-bin state detections as the phase $\beta$ is varied for four different idler projections $\gamma$, with an integration time of 5~min per data point. In Fig.~S\ref{coin2}a and Fig.~S\ref{coin2}c, $\gamma$ is set to 0 (diagonal polarization $\ket{D}$) for the blue curve and $\pi$ (anti-diagonal polarization $\ket{A}$) for the red curve. In Fig.~S\ref{coin2}b and Fig.~S\ref{coin2}d, $\gamma$ is set to $\frac{\pi}{2}$ (left-circular polarization $\ket{L}$) for the orange curve and $\frac{3\pi}{2}$ (right-circular polarization $\ket{R}$) for the green curve. The four coincidence curves have visibilities $91\pm3\%$, $93\pm3\%$, $89\pm4\%$, and $84\pm4\%$ for $\ket{D}$, $\ket{A}$, $\ket{L}$, and $\ket{R}$ respectively, for an average visibility of $89.3\pm1.7\%$. The single-detection events are nearly constant. This shows a non-local form of interference only visible when the entire system is measured, demonstrating the fidelity of phase measurement. No active phase stabilization was implemented over the experimental run time (12 hours).


\begin{thebibliography}{10}
\expandafter\ifx\csname url\endcsname\relax
  \def\url#1{\texttt{#1}}\fi
\expandafter\ifx\csname urlprefix\endcsname\relax\def\urlprefix{URL }\fi
\providecommand{\bibinfo}[2]{#2}
\providecommand{\eprint}[2][]{\url{#2}}

\bibitem{tittel98}
\bibinfo{author}{Tittel, W.}, \bibinfo{author}{Brendel, J.},
  \bibinfo{author}{Zbinden, H.} \& \bibinfo{author}{Gisin, N.}
\newblock \bibinfo{title}{Violation of bell inequalities by photons more than
  10 km apart}.
\newblock \emph{\bibinfo{journal}{Phys. Rev. Lett.}}
  \textbf{\bibinfo{volume}{81}}, \bibinfo{pages}{3563--3566}
  (\bibinfo{year}{1998}).

\bibitem{brendel99timebin}
\bibinfo{author}{Brendel, J.}, \bibinfo{author}{Gisin, N.},
  \bibinfo{author}{Tittel, W.} \& \bibinfo{author}{Zbinden, H.}
\newblock \bibinfo{title}{Pulsed energy-time entangled twin-photon source for
  quantum communication}.
\newblock \emph{\bibinfo{journal}{Phys. Rev. Lett.}}
  \textbf{\bibinfo{volume}{82}}, \bibinfo{pages}{2594--2597}
  (\bibinfo{year}{1999}).

\bibitem{tittel00qcrypt}
\bibinfo{author}{Tittel, W.}, \bibinfo{author}{Brendel, J.},
  \bibinfo{author}{Zbinden, H.} \& \bibinfo{author}{Gisin, N.}
\newblock \bibinfo{title}{Quantum cryptography using entangled photons in
  energy-time bell states}.
\newblock \emph{\bibinfo{journal}{Phys. Rev. Lett.}}
  \textbf{\bibinfo{volume}{84}}, \bibinfo{pages}{4737--4740}
  (\bibinfo{year}{2000}).

\bibitem{marcikic02timebin}
\bibinfo{author}{Marcikic, I.} \emph{et~al.}
\newblock \bibinfo{title}{Time-bin entangled qubits for quantum communication
  created by femtosecond pulses}.
\newblock \emph{\bibinfo{journal}{Phys. Rev. A}} \textbf{\bibinfo{volume}{66}},
  \bibinfo{pages}{062308} (\bibinfo{year}{2002}).

\bibitem{marcikic2004fiftykm}
\bibinfo{author}{Marcikic, I.} \emph{et~al.}
\newblock \bibinfo{title}{Distribution of time-bin entangled qubits over 50~km
  of optical fiber}.
\newblock \emph{\bibinfo{journal}{Phys. Rev. Lett.}}
  \textbf{\bibinfo{volume}{93}}, \bibinfo{pages}{180502}
  (\bibinfo{year}{2004}).

\bibitem{martin12}
\bibinfo{author}{{Martin}, A.} \emph{et~al.}
\newblock \bibinfo{title}{Cross time-bin photonic entanglement for quantum key
  distribution}.
\newblock \emph{\bibinfo{journal}{Phys. Rev. A}} \textbf{\bibinfo{volume}{87}},
  \bibinfo{pages}{020301} (\bibinfo{year}{2013}).

\bibitem{franson89}
\bibinfo{author}{Franson, J.~D.}
\newblock \bibinfo{title}{Bell inequality for position and time}.
\newblock \emph{\bibinfo{journal}{Phys. Rev. Lett.}}
  \textbf{\bibinfo{volume}{62}}, \bibinfo{pages}{2205--2208}
  (\bibinfo{year}{1989}).

\bibitem{franson91}
\bibinfo{author}{Franson, J.~D.}
\newblock \bibinfo{title}{Two-photon interferometry over large distances}.
\newblock \emph{\bibinfo{journal}{Phys. Rev. A}} \textbf{\bibinfo{volume}{44}},
  \bibinfo{pages}{4552--4555} (\bibinfo{year}{1991}).

\bibitem{hadfield2009single}
\bibinfo{author}{Hadfield, R.}
\newblock \bibinfo{title}{Single-photon detectors for optical quantum
  information applications}.
\newblock \emph{\bibinfo{journal}{Nature Photonics}}
  \textbf{\bibinfo{volume}{3}}, \bibinfo{pages}{696--705}
  (\bibinfo{year}{2009}).

\bibitem{shah1988ultrafast}
\bibinfo{author}{Shah, J.}
\newblock \bibinfo{title}{Ultrafast luminescence spectroscopy using sum
  frequency generation}.
\newblock \emph{\bibinfo{journal}{Quantum Electronics, IEEE Journal of}}
  \textbf{\bibinfo{volume}{24}}, \bibinfo{pages}{276--288}
  (\bibinfo{year}{1988}).

\bibitem{dayan2004tpa}
\bibinfo{author}{Dayan, B.}, \bibinfo{author}{Pe'er, A.},
  \bibinfo{author}{Friesem, A.~A.} \& \bibinfo{author}{Silberberg, Y.}
\newblock \bibinfo{title}{Two photon absorption and coherent control with
  broadband down-converted light}.
\newblock \emph{\bibinfo{journal}{Phys. Rev. Lett.}}
  \textbf{\bibinfo{volume}{93}}, \bibinfo{pages}{023005}
  (\bibinfo{year}{2004}).

\bibitem{Huang1992FreqConv}
\bibinfo{author}{Huang, J.} \& \bibinfo{author}{Kumar, P.}
\newblock \bibinfo{title}{Observation of quantum frequency conversion}.
\newblock \emph{\bibinfo{journal}{Phys. Rev. Lett.}}
  \textbf{\bibinfo{volume}{68}}, \bibinfo{pages}{2153--2156}
  (\bibinfo{year}{1992}).

\bibitem{kwiat_upconversion}
\bibinfo{author}{Van{D}evender, A.~P.} \& \bibinfo{author}{Kwiat, P.~G.}
\newblock \bibinfo{title}{High efficiency single photon detection via frequency
  up-conversion}.
\newblock \emph{\bibinfo{journal}{Journal of Modern Optics}}
  \textbf{\bibinfo{volume}{51}}, \bibinfo{pages}{1433--1445}
  (\bibinfo{year}{2004}).

\bibitem{tanzilli2005photonic}
\bibinfo{author}{Tanzilli, S.} \emph{et~al.}
\newblock \bibinfo{title}{A photonic quantum information interface}.
\newblock \emph{\bibinfo{journal}{Nature}} \textbf{\bibinfo{volume}{437}},
  \bibinfo{pages}{116--120} (\bibinfo{year}{2005}).

\bibitem{SFGent_Ramelow_2012}
\bibinfo{author}{Ramelow, S.} \emph{et~al.}
\newblock \bibinfo{title}{Polarization-entanglement-conserving frequency
  conversion of photons}.
\newblock \emph{\bibinfo{journal}{Phys. Rev. A}} \textbf{\bibinfo{volume}{85}},
  \bibinfo{pages}{013845} (\bibinfo{year}{2012}).

\bibitem{kielpinski}
\bibinfo{author}{Kielpinski, D.}, \bibinfo{author}{Corney, J.~F.} \&
  \bibinfo{author}{Wiseman, H.~M.}
\newblock \bibinfo{title}{Quantum optical waveform conversion}.
\newblock \emph{\bibinfo{journal}{Phys. Rev. Lett.}}
  \textbf{\bibinfo{volume}{106}}, \bibinfo{pages}{130501}
  (\bibinfo{year}{2011}).

\bibitem{eckstein2011quantum}
\bibinfo{author}{Eckstein, A.}, \bibinfo{author}{Brecht, B.} \&
  \bibinfo{author}{Silberhorn, C.}
\newblock \bibinfo{title}{A quantum pulse gate based on spectrally engineered
  sum frequency generation.}
\newblock \emph{\bibinfo{journal}{Optics Express}}
  \textbf{\bibinfo{volume}{19}}, \bibinfo{pages}{13770--13778}
  (\bibinfo{year}{2011}).

\bibitem{lavoie12comp}
\bibinfo{author}{Lavoie, J.} \emph{et~al.}
\newblock \bibinfo{title}{Spectral compression of single photons}.
\newblock \emph{\bibinfo{journal}{Nature Photonics}}
  \textbf{\bibinfo{volume}{7}}, \bibinfo{pages}{363--366}
  (\bibinfo{year}{2013}).

\bibitem{deBurghTomo}
\bibinfo{author}{de~Burgh, M.~D.}, \bibinfo{author}{Langford, N.~K.},
  \bibinfo{author}{Doherty, A.~C.} \& \bibinfo{author}{Gilchrist, A.}
\newblock \bibinfo{title}{Choice of measurement sets in qubit tomography}.
\newblock \emph{\bibinfo{journal}{Phys. Rev. A}} \textbf{\bibinfo{volume}{78}},
  \bibinfo{pages}{052122} (\bibinfo{year}{2008}).

\bibitem{takesue2009tomo}
\bibinfo{author}{Takesue, H.} \& \bibinfo{author}{Noguchi, Y.}
\newblock \bibinfo{title}{Implementation of quantum state tomography for
  time-bin entangled photon pairs}.
\newblock \emph{\bibinfo{journal}{Optics Express}}
  \textbf{\bibinfo{volume}{17}}, \bibinfo{pages}{10976--10989}
  (\bibinfo{year}{2009}).

\bibitem{Wang12tomo}
\bibinfo{author}{Wang, S.~X.} \emph{et~al.}
\newblock \bibinfo{title}{High-speed tomography of time-bin-entangled photons
  using a single-measurement setting}.
\newblock \emph{\bibinfo{journal}{Phys. Rev. A}} \textbf{\bibinfo{volume}{86}},
  \bibinfo{pages}{042122} (\bibinfo{year}{2012}).

\bibitem{bell64}
\bibinfo{author}{Bell, J.}
\newblock \bibinfo{title}{On the {E}instein-{P}odolsky-{R}osen paradox}.
\newblock \emph{\bibinfo{journal}{Physics}} \textbf{\bibinfo{volume}{1}},
  \bibinfo{pages}{195--200} (\bibinfo{year}{1964}).

\bibitem{CHSH}
\bibinfo{author}{Clauser, J.~F.}, \bibinfo{author}{Horne, M.~A.},
  \bibinfo{author}{Shimony, A.} \& \bibinfo{author}{Holt, R.~A.}
\newblock \bibinfo{title}{Proposed experiment to test local hidden-variable
  theories}.
\newblock \emph{\bibinfo{journal}{Phys. Rev. Lett.}}
  \textbf{\bibinfo{volume}{23}}, \bibinfo{pages}{880--884}
  (\bibinfo{year}{1969}).

\bibitem{BC_oppositechirp_1}
\bibinfo{author}{Raoult, F.} \emph{et~al.}
\newblock \bibinfo{title}{Efficient generation of narrow-bandwidth picosecond
  pulses by frequency doubling of femtosecond chirped pulses}.
\newblock \emph{\bibinfo{journal}{Optics Letters}}
  \textbf{\bibinfo{volume}{23}}, \bibinfo{pages}{1117--1119}
  (\bibinfo{year}{1998}).

\bibitem{BC_oppositechirp_2}
\bibinfo{author}{Osvay, K.} \& \bibinfo{author}{Ross, I.~N.}
\newblock \bibinfo{title}{Efficient tuneable bandwidth frequency mixing using
  chirped pulses}.
\newblock \emph{\bibinfo{journal}{Opt. Comm.}} \textbf{\bibinfo{volume}{166}},
  \bibinfo{pages}{113--119} (\bibinfo{year}{1999}).

\bibitem{kwiat99spdc}
\bibinfo{author}{Kwiat, P.~G.} \emph{et~al.}
\newblock \bibinfo{title}{Ultrabright source of polarization-entangled
  photons}.
\newblock \emph{\bibinfo{journal}{Phys. Rev. A}} \textbf{\bibinfo{volume}{60}},
  \bibinfo{pages}{R773--R776} (\bibinfo{year}{1999}).

\bibitem{lavoie2009experimental}
\bibinfo{author}{Lavoie, J.}, \bibinfo{author}{Kaltenbaek, R.} \&
  \bibinfo{author}{Resch, K.}
\newblock \bibinfo{title}{Experimental violation of svetlichny's inequality}.
\newblock \emph{\bibinfo{journal}{New Journal of Physics}}
  \textbf{\bibinfo{volume}{11}}, \bibinfo{pages}{073051}
  (\bibinfo{year}{2009}).

\bibitem{mazurek2013dispersion}
\bibinfo{author}{Mazurek, M.~D.} \emph{et~al.}
\newblock \bibinfo{title}{Dispersion-cancelled biological imaging with
  quantum-inspired interferometry}.
\newblock \emph{\bibinfo{journal}{Scientific Reports}}
  \textbf{\bibinfo{volume}{3}}, \bibinfo{pages}{1582} (\bibinfo{year}{2013}).

\bibitem{treacy1969}
\bibinfo{author}{Treacy, E.}
\newblock \bibinfo{title}{Optical pulse compression with diffraction gratings}.
\newblock \emph{\bibinfo{journal}{Quantum Electronics, IEEE Journal of}}
  \textbf{\bibinfo{volume}{5}}, \bibinfo{pages}{454--458}
  (\bibinfo{year}{1969}).

\bibitem{Jezek2003tomo}
\bibinfo{author}{Je\ifmmode~\check{z}\else \v{z}\fi{}ek, M.},
  \bibinfo{author}{Fiur\'a\ifmmode~\check{s}\else \v{s}\fi{}ek, J.} \&
  \bibinfo{author}{Hradil, Z.}
\newblock \bibinfo{title}{Quantum inference of states and processes}.
\newblock \emph{\bibinfo{journal}{Phys. Rev. A}} \textbf{\bibinfo{volume}{68}},
  \bibinfo{pages}{012305} (\bibinfo{year}{2003}).

\bibitem{sensarn2009}
\bibinfo{author}{Sensarn, S.}, \bibinfo{author}{Yin, G.~Y.} \&
  \bibinfo{author}{Harris, S.~E.}
\newblock \bibinfo{title}{Generation and compression of chirped biphotons}.
\newblock \emph{\bibinfo{journal}{Phys. Rev. Lett.}}
  \textbf{\bibinfo{volume}{104}}, \bibinfo{pages}{253602}
  (\bibinfo{year}{2010}).

\end{thebibliography}
\end{document}